\documentclass[preprint,12pt]{elsarticle}
\usepackage{tikz}

\usepackage{color}
\usepackage{mathrsfs}
\usepackage{mathtools}
\usepackage{amsmath}
\usepackage{amssymb}
\usepackage{xcolor}
\usepackage{algorithm}
\usepackage[noend]{algpseudocode}
\usepackage{arydshln}
\usepackage[most]{tcolorbox}
\usepackage{graphicx}
\usepackage{algpseudocode}
\usepackage{caption}
\usepackage[colorlinks = true,
linkcolor = blue,
urlcolor  = blue,
citecolor = blue,
anchorcolor = blue]{hyperref}
\newproof{pf}{Proof}

\newtheorem{assum}{Assumption}

\newtheorem{defn}{Definition}

\newtheorem{thm}{Theorem}
\newtheorem{remark}{Remark}
\newtheorem{corollary}{Corollary}

\global\long\def\R{\mathbb{R}}%
\global\long\def\mod{\mathrm{mod}}
\global\long\def\vec{\mathrm{vec}}%
\global\long\def\vecinverse{\mathrm{vec}^{-1}}%
\global\long\def\diag{\mathop{\mathrm{diag}}}%
\global\long\def\ln{\mathop{\mathrm{ln}}}
\global\long\def\argmin{\mathop{\mathrm{argmin}}}

\usepackage{amssymb}

\begin{document}

\begin{frontmatter}


\title{On input-to-state stability verification of identified models obtained
by Koopman operator}


\author[inst1]{Wenjie Mei}
\author[inst2]{Dongzhe Zheng}
\author[inst3]{Yu Zhou}
\author[inst4]{Ahmad Taha}
\author[inst5]{Chengyan Zhao}

\affiliation[inst1]{organization={School of Automation and Key Laboratory of MCCSE of the Ministry of Education, Southeast University},
            city={Nanjing},
            postcode={210096}, 
            country={China}}
\affiliation[inst2]{organization={Department of Computer Science, Shanghai Jiao Tong University},
            city={Shanghai},
            postcode={200240}, 
            country={China}}
\affiliation[inst3]{organization={Inria, Centrale  Lille, and CRIStAL UMR 9189 CNRS},
            city={Lille},
            postcode={59000}, 
            country={France}}  
\affiliation[inst4]{organization={Civil and Environmental Engineering and Electrical Department, Vanderbilt University},
            city={Nashville},
            postcode={37235}, 
            country={USA}}  
\affiliation[inst5]{organization={Graduate School of Science and Engineering, Ritsumeikan University},
            city={Shiga},
            postcode={525-8577}, 
            country={Japan}}        

\begin{abstract}
This paper proposes a class of basis functions for realizing the input-to-state stability verification of identified models obtained from the true system (assumed to be input-to-state stable) using the Koopman operator. The formulated input-to-state stability conditions are in the form of linear matrix inequalities. Two extensions are presented to relax the imposed restrictions on the basis functions.  Several numerical examples are provided to demonstrate the efficacy of the proposed results.
\end{abstract}

\end{frontmatter}


\section{Introduction}
It has become more facile to collect or measure data from networks/dynamics in recent years, making data-driven methods popular tools in technical systems. For nonlinear control applications, the identification problems may become intractable and, indeed, complex for implementation, especially under significant noise or uncertainty. To that end, a theoretic framework called \emph{Koopman Operator} \cite{koopman1931hamiltonian} has been exploited for lifting complex nonlinear dynamical systems to linear models with higher dimensions, which can be utilized in, for instance, prediction, control, and stability analysis. 

Such a framework is frequently combined with data-driven methods. That is, the collected data from the system trajectory can be used to approximate the Koopman operator, making the application of the Koopman operator more viable. Various related investigations can be found in the directions of its spectral properties \cite{Mezic2005}, linear predictors \cite{Korda2018}, and neural network training \cite{dogra2020optimizing}, to mention a few examples. 

In this work, under the realized system identification (here, system identification \cite{zadeh1962circuit} is defined as determining a system model within a specified class of dynamics based on the measured input/output data) of nonlinear systems (they can be well approximated by “quasi generalized Persidskii systems” \cite{Efimov2019a}), we focus on the stability verification problem in the identified model derived by the Koopman operator theory, as well as its conjunction with a useful data-driven method: the \emph{Extended Dynamic
Mode Decomposition (EDMD)} \cite{Williams2015}.  One can find that there are studies involved in system identification in the context of the Koopman operator in recent years, such as the adaption of Koopman operator theory to identification \cite{mauroy2016linear} and its application to soft robots \cite{Bruder2019};  the investigation on the convergence of a variant of the generator EDMD algorithm \cite{zhang2023quantitative} for calculating a linear representation of the operator. Nevertheless, there still exist gaps in the stability verification issue in identified models (the true system is assumed to be stable) under the employment of the Koopman operator theory. 

For that purpose, this paper proposes a general scheme to examine if an identified system modeled via the Koopman operator technique is input-to-state stable. To the best of the authors' knowledge, rare studies have addressed the stability analysis problem of identified systems obtained using the Koopman operator theory. That is, in practice, the true system is usually assumed to be input-to-state stable. However, due to many factors, such as noise and modeling errors, the input-to-state stability (ISS) of identified models can not be guaranteed. This motivated the works of stability guarantees presented in, for example, \cite{lacy2003subspace} (this work gives a constrained optimization method to guarantee asymptotic stability of identified models derived by subspace identification methods for system identification) and this paper.   On the other hand, stability guarantees for control synthesis are important -- see references \cite{he2014quasi,song2022event,ryu2004stability}

 Due to the intricate forms and possibly highly dynamic nature of nonlinear systems, especially in the presence of external perturbations, the stability analyses can be difficult \cite{Vidyasagar:81:Springer}. Among them, the most widely utilized framework is the ISS concept \cite{sontag1996new,Sontag1995}. 
 As stated above, the stability analysis of an identified model could also be tricky since it can be a nonlinear system taking an unpredictable, thus probably complex form. 
However, the ISS of the true system is necessary for robustness and makes system identification practical~\cite{lacy2003subspace}. These inspire us to deal with the problem of ISS guarantees of identified models. 

The \textbf{main contributions of this work} can be summarized as follows:

    i) If one selects basis functions for the identification as in complex forms, even if only one of them does, the ISS analysis of the resulting identified system (for example, with many nonlinearities) can still be laborious. To that end, we introduce a class of basis functions (they can be linear or nonlinear in generalized Persidskii dynamics) to facilitate the relevant stability analysis since it serves the adaptable number of nonlinearities under the preservation of the generality of the functions. 
    
    ii) Since it is also beneficial to relax the imposed conditions on the basis function, we present two extensions from the considered kind of functions for enlarging the selection range. By doing this, the analyzable forms of the identified system can be further expanded. 
    
    iii) Under a mild assumption (see Assumption~\ref{assum:grey_box} in this paper), we bridge the considered class of nonlinear systems with the actual system. This enables us to skip the obstruction in approximating the differentiation of the input but only focus on being involved in identifying the vector field of the true system.

The rest of this work is organized as follows. 
The lifting approach, the data-driven method for calculating Koopman operators, and the class of nonlinear systems under consideration are  all provided in Section~\ref{sec:Problem}. The considered problem is given in Section~\ref{sec:pro_stat}. Section~\ref{sec:main_results} presents the ISS conditions of identified
systems, followed by introducing two extending forms of the basis functions.  In Section~\ref{sec:example}, we show an example to illustrate
the efficiency of the proposed results. The notation is provided next:

\section*{Notation} 
{\small
\bgroup
\def\arraystretch{1.5}
\begin{tabular}{p{1in}p{3.8in}}
$\displaystyle \mathbb{N}$ &  The set of natural numbers \\
$\displaystyle \mathbb{Z}$ &  The set of integers \\
$\displaystyle \mathbb{R}$ &  The set of real numbers \\
$\displaystyle \R_{+}$ &  The set $\left\{ s\in\R \mid s\geq0\right\}$ \\
$\displaystyle \mathbb{R}^{n}$ &  The vector space of $n$-tuples of real numbers \\
$\displaystyle \mathbb{R}^{m\times n}$ &  The vector space of  $m \times n$ real matrices \\ 
$\displaystyle \rVert\cdot\rVert$ &  The Euclidean norm on
  $\mathbb{R}^{n}$  \\  
  $\displaystyle \rVert A \rVert$ &  The induced matrix norm for a matrix $A$  \\  
    $\displaystyle \rVert A \rVert_F$ &  The Frobenius norm for a matrix $A$  \\ 
    $\displaystyle \mathbb{D}^{n}$ &  The set of $n \times n$ diagonal matrices  \\ 
    $\displaystyle \mathbb{D}^{n}_+$ &  The set of $n \times n$ diagonal matrices with nonnegative diagonal elements \\ 
    $\displaystyle O_{a \times b}$ &  The $a \times b$ zero matrix \\ 
    $\displaystyle \overline{p,n}$ &  The set $\{p,p+1,\dots,n\}$, where $p, n \in\mathbb{N}$ \\ 
  $\displaystyle \tilde{p} \; \mod \; \tilde{n}$  &  The remainder of the Euclidean division of $\tilde{p}$ by $\tilde{n}$, where $\tilde{p}, \tilde{n} \in\mathbb{N}$ \\ 
    $\displaystyle \vec(A)$ &  The vectorization of a matrix $A^{m \times n}$, that is, for a vector $\ell \in \mathbb{R}^{mn}$, $\vecinverse_{m \times n}(\ell)=(\vec(I_n)^\top \otimes I_m)(I_n \otimes \ell) \in \mathbb{R}^{m \times n}$, where $I_n$ denotes the $n \times n$ identity matrix and $\otimes$ denotes the  Kronecker product \\ 
    $\displaystyle \lfloor  \cdot \rfloor$ &  The floor function defined on $\R$ \\ 
    $\displaystyle C(U,R)$ &  The space of continuous functions $f\colon U \rightarrow R$, where $U, R$ are metric spaces  \\ 
    $\displaystyle C^{1}_{n}([t_{1},t_{2}))$ &  The Banach space of continuously differentiable
  functions $\psi \colon [t_{1},t_{2}) \to\mathbb{R}^{n}$ with the norm
  $\Vert\psi\Vert_{[t_{1},t_{2})}=\sup_{ r \in [t_1,t_2)} \|\psi(r)\| + \sup_{ r \in [t_1,t_2)}\|\frac{d \psi(r)}{d r}\| < +\infty$  \\ 
   $\displaystyle \mathscr{K}$ &  A continuous function class: Function $\sigma \colon \mathbb{R}_{+}\to\mathbb{R}_{+}$ is strictly increasing and $\sigma(0)=0$ \\ 
  $\displaystyle \mathscr{K}_\infty$ &  A continuous function class: Function $\sigma \colon \mathbb{R}_{+}\to\mathbb{R}_{+}$ is strictly increasing, $\sigma(0)=0$, and $\lim_{r \to \infty} \sigma(r) = \infty$ \\
  $\displaystyle \mathscr{KL}$ &  A continuous function class: Function $\beta \colon \mathbb{R}_{+} \times \mathbb{R}_{+} \to\mathbb{R}_{+}$, $\beta(\cdot,t) \in \mathscr{K}$ for each fixed $t \in \R_{+}$ and $\beta(s,\cdot)$ is decreasing to zero for each fixed $s \in \R_{+}$
\end{tabular}
\egroup }

\vspace{0.5cm}

\section{Preliminaries} \label{sec:Problem}
This section introduces the lifting approach based on utilizing the Koopman operator, which can be represented as a matrix after projection. Based on that, the relationship between the Koopman operator and system identification is clarified. Then, the EDMD algorithm (see \emph{e.g.}, \cite{Williams2015}) is shown for approximating the Koopman operator so that a system identification can be realized. 

\subsection{Representing Koopman operator via a matrix}
Following the system definitions given in the~\ref{appendix:general_system}, for brevity, consider equivalently rewriting system~\eqref{eq:general_nonlinear_system} to the dynamics 
\begin{gather} \label{eq:extended_general_nonlinear_system}
\dot{\chi}(t) = \tilde{F}(\chi(t)):= \begin{bmatrix}  F(x(t),u(t)) \\ \ \dot{u}(t) \end{bmatrix}, \quad t \in \mathbb{R}_+
\end{gather}
with an extended state $\chi = \begin{bmatrix} x^\top & u^\top \end{bmatrix}^\top \in \mathbb{R}^{n+m}$, which admits a unique solution $\chi^t(\chi_0)$ for any initial condition $\chi_0\in \mathbb{R}^{n+m}$ defined for $t\in[0,T_{\chi_0})$ with some $T_{\chi_0}>0$. Then, the Koopman operator theory states that
\[
K^t \tilde{H}  = \tilde{H} \circ \chi^t,
\]
where $\tilde{H} \colon \R^n \times \R^m \rightarrow \R$ are observable functions in a space $\tilde{\mathcal{H}}$ and $K^t$ is the Koopman operator. In order to define a practical procedure for lifting the dimension of the model \eqref{eq:extended_general_nonlinear_system} for using the data-driven method that will be presented in the sequel, we need to introduce a finite-dimensional subspace of $\tilde{\mathcal{H}}$ as follows: 
$
\tilde{\mathcal{H}}^{N_{H}} := \text{span}\;\{H_1(\chi) \dots H_{N_{H}}(\chi) \} 
$
spanned by $N_{H} $ (a finite integer) linear independent scalar-valued functions $H_1, \dots, H_{N_{H}}$, which are called \emph{Lifting Functions} \cite{Bruder2019}. Then,  the projection of $\tilde{H}$ onto the space $\tilde{\mathcal{H}}^{N_{H}} (\subset \tilde{\mathcal{H}})$ can be expressed as:
$$
\mathbb{P} \; \tilde{H}(\chi) = a^\top {\small \begin{bmatrix}  H_1(\chi) \\ \vdots \\ H_{N_{H}}(\chi) \end{bmatrix}} := a^\top P(\chi)  \in \mathbb{R}, \quad a \in \R^{N_{H}}, 
$$
and
\begin{gather} \label{eq:projected_space_N_H}
\mathbb{P} \; K^t \tilde{H}(\chi) = b^\top P(\chi)  \in \mathbb{R}, \quad b \in \R^{N_{H}}, 
\end{gather}
where $P(\chi) = \left[H_1(\chi)  \dots  H_{N_H}(\chi) \right]^\top$ and $\mathbb{P}:\tilde{\mathcal{H}} \rightarrow  \tilde{\mathcal{H}}^{N_{H}} $ is a projection operator. To obtain \eqref{eq:projected_space_N_H}, we use a property of the operator $\mathbb{P}$, \emph{i.e.}, $K^t: \tilde{\mathcal{H}}^{N_H} \to \tilde{\mathcal{H}}$, such that
$
  K^t_{\text{rep}} a =  b,
$
where the matrix $K^t_{\text{rep}}$ is a representation of $K^t$ \cite{mauroy2016linear}.

In short, the matrix $K^t_{\text{rep}}$ is a linear representation of the nonlinear dynamic system~\eqref{eq:extended_general_nonlinear_system}. It can be calculated numerically using data-driven approaches, such as the EDMD, which will be briefly introduced later.  Notice that if one selects lifting functions under the restriction of \emph{$K^t$-invariance}, \emph{i.e.}, $K^t (\tilde{\mathcal{H}}^{N_H}) = \tilde{\mathcal{H}}^{N_H}$, then in~\eqref{eq:projected_space_N_H} the projection operator $\mathbb{P}$ can be omitted.

\vspace{-0.2cm}

\subsection{Connecting Koopman operator and system identification}

In this subsection, the relationship between the Koopman operator and system identification \cite{mauroy2016linear} is clarified, illustrating how to identify the vector field $\tilde{F}$ by applying the Koopman operator technique. 

For system~\eqref{eq:extended_general_nonlinear_system},  assume that the function $\tilde{F}_i$ can be projected onto the space $\tilde{\mathcal{H}}^{N_{F}}$ (here $\tilde{\mathcal{H}}^{N_{F}} := \text{span}\;\{G_1(\chi) \dots G_{N_{F}}(\chi) \}$ with the linear independent \emph{Basis Functions} $G_1 \dots G_{N_{F}}$; note that $N_F$ is not necessarily equal to $N_H$) as
\begin{gather} \label{eq:vector_filed_projection}
\tilde{F}_{i}(\chi) =  \sum_{j=1}^{N_{F}} \lambda_{ij} G_j(\chi), \quad \forall i \in \overline{1,n+m},
\end{gather}
where $\tilde{F}_{i}$ is the $i$-th element of the function $\tilde{F}$ defined in~\eqref{eq:extended_general_nonlinear_system} and the specific forms of basis functions are selected by a designer in practice. By the Koopman operator theory, there is a Koopman infinitesimal generator $L=\sum_{i=1}^{n+m} \sum_{j=1}^{N_{F}} \lambda_{ij}  L_{ij}$ with the operators $L_{ij} = G_j  \cdot \frac{\partial}{\partial \chi_i}$ such that  
\begin{gather*}
\begin{aligned}
\dot{P}(\chi)  &=  \sum_{i=1}^{n+m} \dot{\chi}_i \cdot \frac{\partial P(\chi)}{\partial \chi_i} = \sum_{i=1}^{n+m} \tilde{F}_i \cdot \frac{\partial P(\chi)}{\partial \chi_i}  \\
& =  \sum_{i=1}^{n+m} \sum_{j=1}^{N_{F}} \left(\lambda_{ij} G_j  \right) \cdot \frac{\partial P(\chi)}{\partial \chi_i}   =  \sum_{i=1}^{n+m} \sum_{j=1}^{N_{F}} \lambda_{ij} \left(G_j  \cdot \frac{\partial P(\chi)}{\partial \chi_i}\right)  \\ 
& = LP(\chi),
\end{aligned}
\end{gather*}
which is a linear system so that 
\begin{equation}\label{eq:linear_system_KP}
\boxed{P(\chi(T+ t))  = e^{L t} P(\chi(T)) = K^{t} P(\chi(T)),}
\end{equation}
where $T >0$. 

Until now, we have introduced the linear operator $K^t$ and demonstrated the connections among the Koopman operator $K^t$, the generators $L_{ij}$ and the identification of $\tilde{F}$, for the implementation of the system identification procedures, the forms of $G_j$ should be properly selected by a designer. The remaining part of this section provides a way to calculate the values of $\lambda_{ij}$ to actualize a system identification.

Note that the operator $L$ can also be represented by a matrix 
\begin{gather}
L_{\text{rep}} = \sum_{i=1}^{n+m} \sum_{j=1}^{N_{F}} \lambda_{ij}  L_{ij,\;\text{rep}}, \label{L_rep_sum}
\end{gather}
which can be computed by  
\begin{gather} \label{eq:log_K}
L_{\text{rep}} = \frac{1}{t} \ln(K^t_{\text{rep}})
\end{gather}
due to the relation~\eqref{eq:linear_system_KP}. 
In addition, by the relation~\eqref{L_rep_sum}, it can be deduced that
\begin{align} \label{eq:Gamma_cal}
\Gamma &:= \begin{bmatrix} \lambda_{11} & \dots & \lambda_{1N_F} & \dots & \lambda_{(m+n)1} & \dots & \lambda_{(m+n)N_F}  \end{bmatrix}^\top  \\ 
&=  \left[ \begin{matrix} \vec(L_{11,\;\text{rep}}) & \dots & \vec(L_{1N_F,\;\text{rep}}) & \dots\end{matrix}  \right. \nonumber   \\  
&\left. \begin{matrix} \vec(L_{(n+m)1,\;\text{rep}})  &\dots  &  \vec(L_{(n+m)N_F,\;\text{rep}}) \end{matrix} \right]^{\dagger} \vec(L_{\text{rep}}), \nonumber
\end{align}
where the symbol $\dagger$ stands for the pseudoinverse. We will describe how $K^t_{\text{rep}}, L_{\text{rep}}$ can be obtained by the EDMD algorithm in the next section. Thus, if one has the values of $L_{ij,\;\text{rep}}$, then it is direct to derive the parameters $\lambda_{ij}$ via~\eqref{eq:Gamma_cal}.


It is worth mentioning here that, in practice preserving the $K^t$-invariance is challenging since each operator $L_{ij}$ generates terms that may increase the amount of linear independent lifting functions in the resulting space (in practice (the reader is referred to \eqref{eq:projected_space_N_H} for illustration), the Koopman operator $K^t: \tilde{\mathcal{H}}^{N_H} \to \tilde{\mathcal{H}}$ and $\tilde{\mathcal{H}}^{N_H} \subset \tilde{\mathcal{H}}$, which is different from the definition: $K^t: \tilde{\mathcal{H}} \to \tilde{\mathcal{H}}$ in theory). Then, the insight of this work is that by selecting $G_j$ and the elements $P_l$ of $P$ satisfied the sector boundedness condition formulated in Assumption~\ref{main_assum_sector_condition} of~\ref{appendix:gps}, 
there always exists an operator $L_{ij}$ such that 
$
L_{ij} P = G_j \cdot \frac{\partial P}{\partial \chi_i}
$
since the scalar-valued functions $G_j, P_l$ are in the semigroup  $\left(\mathcal{F}(\mathbb{R}),\cdot \right)$, where $\mathcal{F}(\mathbb{R}) = \{ f \in C(\mathbb{R}, \mathbb{R})  \mid  f(0)= 0, \nu f(\nu)>0, \; \forall \; \nu \neq 0 \}$.
This provides a possible scheme to ensure the \emph{$K^t$-invariance} in theory, which demonstrates one of the strengths of the criterion of choosing $G_j$ as functions  satisfying Assumption~\ref{main_assum_sector_condition}, whose further merits will be interpreted at length in the sequel.

\subsection{Approximating Koopman operator}
This section presents a data-driven approach: the EDMD algorithm (see \cite{Williams2015} for reference). It can calculate a matrix $K_{\text{rep}}$ (for brevity, the variable $t$ in $K^t_{\text{rep}}$ is omitted, with the understanding that it is implicitly included) to represent the Koopman operator.


For a given constant sampling time $T_c  >0$ and a series of the input values $ \{u_{k}\}_{k =1}^{N+1}$, where the length $N\in\mathbb{N}$, collect pairs of the state $x_k$ and the output $y_k$ of the true system: $\{(x_k, u_k),(y_{k},u_{k+1})\}_{k=1}^{N}$ with $u_{k+1}= \Theta u_{k}$,  where $\Theta$ is a left shift operator. Here, $y_k$ is not necessarily equal to $x_{k+1}=x^{T_c}(x_{k},u)$ due to the existence of measurement noises. In the case that there do not exist such noises, one can assume that $y_k = x^{T_c}(x_{k},u)$.

The EDMD converts an approximation problem to a minimization one as follows: 
\begin{gather}
K_{\text{rep}} = \argmin_{K_{\text{rep}}}  \sum_{k=1}^N \left \rVert P(y_{k}, u_{k+1}) - (K_{\text{rep}})^\top P(x_k, u_k) \right \rVert_F. \label{EDMD_minimization}
\end{gather} 
Here, recall that $L_{\text{rep}}$ and $K_{\text{rep}}$ denote the matrix representations of the operators $L$ and $K$, respectively. Then, the corresponding least square solution of~\eqref{EDMD_minimization} can be dervied by
\begin{gather} \label{least_square_solution}
K_{\text{rep}}  = \begin{bmatrix}  P(x_1, u_1)^\top \\ \vdots \\ P(x_N, u_N)^\top \end{bmatrix}^{\dagger} \begin{bmatrix}  P(y_1, u_2)^\top \\ \vdots \\ P(y_N, u_{N+1})^\top \end{bmatrix}.
\end{gather}


So far, we have introduced the nonlinear system \eqref{eq:extended_general_nonlinear_system} into account to give the general definitions of the lifting approach, the Koopman operator and its connection with identification, and the EDMD algorithm.

\section{Problem statement} \label{sec:pro_stat}

In this paper, we assume that the considered identified models take the form of generalized Persidskii dynamics, and the following assumption is imposed on the true system~\eqref{eq:extended_general_nonlinear_system} (see~\ref{appendix:general_system} for the detailed description of~\eqref{eq:extended_general_nonlinear_system}).

\begin{assum} \label{assum:grey_box}
Assume that the true system~\eqref{eq:extended_general_nonlinear_system} is ISS and can be presented in the form 
\begin{gather} \label{eq:system_for_identification}
\dot{x} = f(x) + Du,
\end{gather}
where $f:\mathbb{R}^n \to \mathbb{R}^n$ is a vector field and $D\in\mathbb{R}^{n\times m}$ is a constant matrix. 
\end{assum}


\begin{remark}
In this work, we use generalized Persidskii systems to approximate the true system~\eqref{eq:system_for_identification}. Then, one can see that non-restrictive (since~\eqref{eq:system_for_identification} is also a general nonlinear system that has been widely studied) Assumption~\ref{assum:grey_box}  makes the identification feasible since the continuous function $f$  can be approximated (or identified) arbitrarily well by a neural network represented by the generalized Persidskii dynamics (see \cite{mei2024annular} for the details of representing recurrent neural networks by generalized Persidskii systems; also, see Universal Approximation Theorem \cite{hornik1991approximation,bullinaria2013recurrent} for the illustration that a recurrent neural network can approximate the function $f$ arbitrarily well). 

On the other hand, at least the generalized Persidskii system is more adaptable for approximating the true system~\eqref{eq:system_for_identification} than the linearization of~\eqref{eq:system_for_identification} (locally or globally), considering, for example, $\dot{x} = -x\tanh(x) + Du = -x^2 + \frac{x^4}{3} - \frac{2x^6}{15}+ \dots + Du$, and the latter is a particular case of the former one. 
\end{remark}

The main objective of this work is to propose a generic scheme to verify if an identified model obtained by applying the Koopman operator theory is ISS, under the preliminary that the true system~\eqref{eq:system_for_identification} is ISS and its implicit form is known (\emph{i.e.} Assumption~\ref{assum:grey_box} is fulfilled).

\section{Main results: Basis functions for ISS analysis of identified models} \label{sec:main_results}

In this section, we propose a novel type of basis functions (they belong to the nonlinearity classes in generalized Persidskii systems) useful in analyzing the ISS property of an identified model associated with the system~\eqref{eq:system_for_identification}, which is also the main contribution of this study.

\subsection{A class of basis functions} \label{sec:a_proposed_class_basis_functions}
The considered type of basis functions $G_j(\chi)$ (see \eqref{eq:vector_filed_projection} to recall their definition) is defined as follows.
\begin{defn}
The functions $G_j$ satisfying Assumption~\ref{main_assum_sector_condition} are called \emph{Sector Basis Functions (SBFs)}.
\end{defn}

We then show the usefulness of SBFs in system identification. Firstly, to suppress unnecessary computation of the given input $u$ during the solution process of~\eqref{EDMD_minimization}, we further impose that the basis functions take the form
$
G_s (\chi) = \varphi_s(x) + \psi_s(u), \; s \in \overline{1,nM+m},
$
where the functions $\varphi_s : \mathbb{R}^{n} \to \mathbb{R}$ and $\psi_s : \mathrm{U} \to \mathbb{R}$. Moreover, let $\psi_s(u)=0$ for $s \in \overline{1,nM}$ and $G_{s'}(\chi) = u_{s'-nM}$ for $s' \in \overline{nM+1,nM+m}$. We have
\begin{equation}
    G(\chi) = \begin{bmatrix}  G_1(x) \\ \vdots \\  G_{nM+m}(x) \end{bmatrix}   := \begin{bmatrix}  \varphi_1(x) \\ \vdots \\  \varphi_{nM}(x) \\ u \end{bmatrix}.
\end{equation}
Here, $N_F = nM+m$ (see Equation \eqref{eq:vector_filed_projection} for the explanation of $N_F$). Recapture that the $s$-th element of the vector-valued function $G(\chi)$ is a basis function. 

Then, with conciseness we apply scalar-valued SBFs $\tilde{f}_{1},\dots,\tilde{f}_{nM}$ to $G(\chi)$ as:  
\[
f_{j'}(x) =  { \small \begin{bmatrix}  \varphi_{(j'-1)n+1}(x) \\ \vdots \\ \varphi_{j'n}(x) \end{bmatrix}  =  \begin{bmatrix}  \tilde{f}_{(j'-1)n+1}(x_1) \\ \vdots \\ \tilde{f}_{j'n}(x_n) \end{bmatrix}}  \in \R^n
\]
 for all $j' \in \overline{1,M}$, 
resulting in 
\begin{gather} \label{eq:basis_functions_selected}
G(\chi) = \begin{bmatrix}  f_1(x)  \\ \vdots \\ f_M(x)  \\ u \end{bmatrix}.
\end{gather}
where $f_1, \dots, f_M \colon \mathbb{R}^n \to \mathbb{R}^n$ are vector-valued SBFs. The identification of $u$ is not of interest since it is known. Hence, without loss of generality and by \eqref{eq:vector_filed_projection}, we deal solely with identifying the vector field $F$:
\[
\tilde{F}(\chi) = \begin{bmatrix} F(x,u) \\ \dot{u} \end{bmatrix} = \left[
\begin{array}{c;{2pt/2pt}cc}
     \Gamma_1 \; ... \; \Gamma_M  & B \\ \hdashline[2pt/2pt]
      ... & ... 
\end{array}
\right]G(\chi),
\]
leading to
\[
F(x,u) = \sum_{j'=1}^{M} \Gamma_{j'} f_{j'}(x) + B u, 
\]
where { \small
\begin{gather} 
\begin{aligned}
\Gamma_{j'} &= \left(\vecinverse_{n \times n} \left(E_{j'} \Gamma \right) \right)^\top, \; E_{j'} \in \mathbb{R}^{n^2 \times ((m+n)(nM+m))}, \nonumber \\ 
[E_{j'}]_{ab} &= \begin{cases}  1, \; \text{if} \;   b =  \lfloor \frac{a}{n} \rfloor (nM+m) +(j'-1)n+  a \; \mod \; n, \\ \; a \; \mod \; n \neq 0; \\ 
1, \; \text{if} \; b =  (\lfloor \frac{a}{n} \rfloor-1) (nM+m) + j'n, \;  a \; \mod \; n = 0; \\ 
0, \; \text{otherwise}, \end{cases} \nonumber \\
B & = \left(\vecinverse_{m \times n} \left(\tilde{B} \Gamma \right) \right)^\top, \; \tilde{B} \in \R^{(mn) \times ((m+n)(nM+m))}, \nonumber \\ 
[\tilde{B}]_{cd} &= \begin{cases} 1, \; \text{if} \;   d =  \lfloor \frac{c}{m} \rfloor (nM+m) +nM+  c \; \mod \; m, \\ \; c \; \mod \; m \neq 0; \\ 
1, \; \text{if} \; d =  \lfloor \frac{c}{m} \rfloor  (nM+m), \;  c \; \mod \; m = 0;
\\ 0, \; \text{otherwise}. \end{cases} \label{eq:compute_parameters_B_Gamma}
\end{aligned}
\end{gather}}
Here we recall that the vector $\Gamma$ is defined in~\eqref{eq:Gamma_cal} and emphasize that the matrices $\Gamma_{j'}$ and $B$ are essentially the extraction and reorganization of the elements (the parameters $\lambda_{ij}$) in $\Gamma$, \emph{i.e.}
\[
   \Gamma_{j'} =  {\small \begin{bmatrix} 
\lambda_{1((j'-1)n+1)} & \dots & \lambda_{1(j'n)} \\
\vdots & \ddots & \vdots \\
\lambda_{n((j'-1)n+1)} & \dots & \lambda_{n(j'n)}
\end{bmatrix}},\;
B = {\small \begin{bmatrix} 
\lambda_{1(Mn+1)} & \dots & \lambda_{1(Mn+m)} \\
\vdots & \ddots & \vdots \\
\lambda_{n(Mn+1)} & \dots & \lambda_{n(Mn+m)}
\end{bmatrix}}.
\]
Also, the matrices $E_{j'}, \tilde{B}$ are used for the extractions.  Therefore, the identified system
\begin{gather}
\boxed{\dot{x} = \sum_{j'=1}^{M} \Gamma_{j'} f_{j'}(x) + B u }\label{F_x_approximation}
\end{gather}
is in the form of generalized Persidskii systems~\eqref{eq:main_system_Per} ($f_1,\; \dots,\;f_M$ are SBFs).

The notable advantages of considering the generalized Persidskii system are that its stability properties have been well investigated in, for instance, \cite{mei2021input,Efimov2019a}, and letting basis functions take the form of such a class of dynamics allows us to analyze the stability of the identified model (random selections may lead to complex stability analysis) and preserve the generality of basis functions to some extent. Apart from these, the theoretically infinite number of nonlinearities in the generalized Persidskii systems can be well-fitted into the possibly voluminous dimension of the lifting/basis function.

\subsection{ISS property of the identified model}
In this section, the ISS conditions of the identified system are formulated, based on the selection of SBFs, for verifying the ISS property of \eqref{F_x_approximation} under Assumption~\ref{assum:grey_box}. The following theorem investigates the ISS of the identified model \eqref{F_x_approximation}.

\begin{thm}
\label{thm:main_ISS} Let Assumption \ref{main_assum_sector_condition}  be satisfied with $\phi \in \mathbb{N} \backslash \{0\}$.
If there exist $0\leq P=P^{\top}\in\mathbb{R}^{n\times n}$; $\left\{\Lambda^{j}=\diag(\Lambda_{1}^{j},\dots,\Lambda_{n}^{j}) \right\}_{j=1}^{M}$, $\left\{\Xi^{k} \right\}_{k=1}^{M}$, $\left\{ \Upsilon_{s,z} \right\}_{0 \leq s < z \leq M}$ $\subset \mathbb{D}_{+}^{n}$; $0<\Phi=\Phi^{\top}\in\mathbb{R}^{n\times n}$; $\rho \in \R$
such that 
\begin{equation} \label{eq:main_thm_LMIs}
 P + \rho \sum_{j=1}^{\mu}\Lambda^{j}>0, \quad Q \leq 0, \quad
\sum_{k=1}^{\phi} \Xi^k + 2 \sum_{s=0}^{\phi} \sum_{z=s+1}^{\phi} \Upsilon_{s,z} >0,
\end{equation}
where {\scriptsize
\begin{gather*}
Q_{1,1}=O_{n \times n}; \quad Q_{j+1,j+1}= \Gamma_{j}^{\top}\Lambda^{j}+\Lambda^{j} \Gamma_{j}+\Xi^{j},\;j\in\overline{1,M} \\
Q_{1,j+1}=P \Gamma_{j} +\Upsilon_{0,j},\;j\in\overline{1,M}; \quad Q_{1,M+2} = P, \\
Q_{s+1,z+1}= \Gamma_{s}^{\top}\Lambda^{z}+\Lambda^{s} \Gamma_{z}+\Upsilon_{s,z},\;s\in\overline{1,M-1},\;z\in\overline{s+1,M}, \\
Q_{j+1,M+2} = \Lambda^j; \quad Q_{M+2,M+2} = -\Phi,
\end{gather*}}  
and $\phi$ is defined in~\ref{appendix:gps}, then a forward complete system \eqref{F_x_approximation} is ISS. 
\end{thm}
\begin{pf}
A similar proof can be founded in \cite{mei2021input,Efimov2019a,mei2022delay}, where the ISS
analysis of \eqref{F_x_approximation} can be performed using a Lyapunov function
$
V(x)=x^\top Px+2\sum_{j \in \overline{1,M}}\sum_{i \in \overline{1,n}}\Lambda^j_i \int_{0}^{ x_i} f_j^i(\nu)d \nu,
$
where $f^i_j$ is the $i$-th element of $f_j$, and $0 \leq P = P^\top\in\R^{n\times n}$ and
$\Lambda^j=\diag(\Lambda^j_1,...,\Lambda^j_n) \in \mathbb{D}_+^{n}$ are
tuning matrices. 

By  Finsler’s Lemma, the first condition in~\eqref{eq:main_thm_LMIs} ensures that \(x^\top Px > 0\) on the subspace where \(\sum_{j=1}^{\mu}\Lambda^{j} = 0\), provided \(x \neq 0\) (it is important to note that under Assumption 3, the integral $\int_{0}^{x_i} f_j^i(\nu)\,d\nu$
equals zero only when \(x_i = 0\); otherwise, it is strictly positive).
This guarantees positive
definiteness of $V$ under Assumption~\ref{main_assum_sector_condition}. Then, there exist functions
$\alpha_1^{P,\Lambda^1,\dots,\Lambda^M},
\alpha_2^{P,\Lambda^1,\dots,\Lambda^M}\in\mathcal{K}_\infty$ such that 
\[
\alpha_1^{P,\Lambda^1,\dots,\Lambda^M}(\|x\|)\leq V(x)\leq\alpha_2^{P,\Lambda^1,\dots,\Lambda^M}(\|x\|)
\] for all $x\in \R^n$. For instance, the function $\alpha_2^{P,\Lambda^1,\dots,\Lambda^M}$ can always be taken as {\scriptsize
\begin{gather*} 
\alpha_2^{P,\Lambda^1,\dots,\Lambda^M}(\tau) = \lambda_{\max}(P)
\tau^2 + 2 Mn  \max_{j \in
  \overline{1,M}, i \in \overline{1,n}} \left\{ \Lambda^j_i
\int_{0}^{\tau} f_j^i(\nu) \ d\nu \right\}.
\end{gather*} }
These recover the first relation in~\eqref{eq:ISS_V}.  

Next, consider the second condition in~\eqref{eq:ISS_V}: Taking the time derivative $\dot{V} = \nabla V(x)\dot{x}$, we have {\small
\begin{gather*}
\begin{aligned}
\dot{V} =&  { \scriptsize\left[\begin{array}{c}
x\\
f_{1}(x)\\
\vdots\\
f_{M}(x)\\
Bu
\end{array}\right]^{\top} } Q {\scriptsize \left[\begin{array}{c}
x\\
f_{1}(x)\\
\vdots\\
f_{M}(x)\\
Bu
\end{array}\right]}
 -\sum_{j=1}^{M}f_{j}(x)^{\top}\Xi^{j}f_{j}(x) 
 -2\sum_{j=1}^{M}x^{\top}\Upsilon_{0,j}f_{j}(x)
 \\ 
 &-2\sum_{s=1}^{M-1}\sum_{z=s+1}^{M}f_{s}(x)^{\top}\Upsilon_{s,z}f_{z}(x)+(Bu)^{\top}\Phi Bu \\
 \leq  & -\sum_{j=1}^{M}f_{j}(x)^{\top}\Xi^{j}f_{j}(x)-2\sum_{j=1}^{M}x^{\top}\Upsilon_{0,j}f_{j}(x) \\
 &-2\sum_{s=1}^{M-1}\sum_{z=s+1}^{M}f_{s}(x)^{\top}\Upsilon_{s,z}f_{z}(x)+(Bu)^{\top}\Phi Bu.   
 \end{aligned}
\end{gather*}}
If the last LMI of~\eqref{eq:main_thm_LMIs} holds (with all matrices being positive semi-definite), then for the $i$-th element of diagonal matrices $\Xi^k, \Upsilon_{s,z}$ (here $k \in \overline{1,\phi}$, $s \in \overline{0,\phi}$, $z \in \overline{s+1,\phi}$), at least one of them (the $i$-th elements) is positive. Consequently, the total term \(-\sum_{j=1}^{M} f_j(x)^{\top} \Xi^j f_j(x) - 2\sum_{j=1}^{M} x^{\top} \Upsilon_{0,j} f_j(x) 
- 2\sum_{s=1}^{M-1} \sum_{z=s+1}^{M} f_s(x)^{\top} \Upsilon_{s,z} f_z(x)\) is radially unbounded, ensuring that the second condition in~\eqref{eq:ISS_V} of the revision is satisfied. Note that only the nonlinearities $f_1,\dots,f_{\phi}$ are radially unbounded (refer to the remarks following Assumption~\ref{main_assum_sector_condition} for further clarification). By Theorem~\ref{thm:main_ISS_Ly}, the proof is complete. 

\end{pf}

From Section~\ref{sec:a_proposed_class_basis_functions} to Theorem~\ref{thm:main_ISS}, we have placed mild restrictions on a problem with great freedom to obtain an analyzable one (in terms of ISS analysis). In addition, the significance of stability verification of identified systems and the retention of the generality of basis functions indicate the acceptable price is worth it. These reflect the main novelty of this work.

\subsection{Extensions of basis functions}

This section demonstrates that extending the considered basis functions is possible to increase the selection range when one implements data-driven methods. Let us consider the first approach of relaxing the imposed form of  basis functions: 

1) \emph{Translation of functions}: We start by defining a vector-valued function  $G(\nu) = \begin{bmatrix} g_1(\nu) & \dots & g_n(\nu) \end{bmatrix}^\top=F(\nu) + \ell = \begin{bmatrix} f_1(\nu) & \dots & f_n(\nu) \end{bmatrix}^\top + \ell$ for all $\nu \in \mathbb{R}^n$, where  $f_1,\dots, f_n \in \mathcal{F}(\mathbb{R})$ and $\ell \in \mathbb{R}^n$. This formulation can also be expressed as: $G(\nu)- \ell = F(\nu)$, where the functions $f_i$ are in the semigroup $\mathcal{F}(\mathbb{R})$. It is clear that $g_i$ do not satisfy Assumption~\ref{main_assum_sector_condition} if the $i$-th element of $\ell$ is not equal to zero. Afterwards, we can extend the system~\eqref{eq:main_system_Per} to
$
\dot{x}(t)=A_{0}x(t)+\sum_{j'=1}^{M}A_{j'}G_{j'}(x(t))+u(t),
$
which is essentially equivalent to 
$
\dot{x}(t)=A_{0}x(t)+\sum_{j'=1}^{M}A_{j'}F_{j'}(x(t))+ u(t) + \sum_{j'=1}^{M}A_{j'} l_{j'},
$
where the constant vectors $l_1,\dots,l_M \in \mathbb{R}^n$. Therefore, one can see that the generality of the functions $G_{j'}$ is greater than that of $F_{j'}$, which illustrates a possible direction of extensions. It is also worth mentioning that such a transformation does not affect the use of the lifting approach and the stability analysis introduced above.

2) \emph{Change of independent variables of functions}: We consider generalizing the variables of the functions $F_{j'}$, for which the linear operators $\mathcal{T}_{j'} x := R_{j'} x$ are taken into account, where $R_{j'} \in \mathbb{R}^{k_{j'} \times n}$ are constant matrices with appropriate dimensions ($\mathcal{T}_{j'}$ may  be the identity operator, then the independent variables do not change). Thus, under the substitutions of $x$ to $\mathcal{T}_{j'} x$ in the nonlinearities of the system~\eqref{eq:main_system_Per}, we can obtain the resulting system:
\vspace{-0.7em}
\begin{equation} \label{eq:change_variable_P_system}
\dot{x}(t)=A_{0}x(t)+\sum_{j'=1}^{M}A_{j'}F_{j'}( R_{j'} x(t))+ u(t),
\end{equation}
extending the model \eqref{eq:main_system_Per}. For such a generalization, we require a minor revision of Assumption~\ref{main_assum_sector_condition}, provided next. 
\begin{assum} \label{assum:modify_main_assum}
Assume that for any $i\in\overline{1,k_{j'}}$
and $j'\in\overline{1,M}$, $\nu f^{i}_{j'}(\nu)>0$, for all
$\nu\neq 0$.
\end{assum}

Given this assumption\footnote{There exists an index $\phi\in\overline{0,M}$ such that
for all $s\in\overline{1,\phi}$ and $i\in\overline{1,k_{s}}$,
$\displaystyle
\lim_{\nu\rightarrow\pm\infty}f^{i}_{s}(\nu)=\pm\infty$. Also, there
exists $\mu\in\overline{\phi,M}$ such that for all
$s\in\overline{1,\mu}$, $i\in\overline{1,k_{s}}$, we have
$\displaystyle
\lim_{\nu\rightarrow\pm\infty}\int_{0}^{\nu}f^{i}_{s}(r)dr=+\infty.$}, we can now formulate the ISS conditions for the system~\eqref{eq:change_variable_P_system} in the following corollary.

\begin{corollary} \label{cor:extension_basis_function}
 Let Assumption \ref{assum:modify_main_assum} be satisfied with $\phi \in \mathbb{N} \backslash \{0\}$. If there exist  $0\leq P =P^{\top}\in\mathbb{R}^{n\times n}$;
$\Lambda^{j}=\mathrm{diag}(\Lambda_{1}^{j},\dots,\Lambda_{k_{j}}^{j})\in
\mathbb{D}_{+}^{k_{j}} \ (j \in \overline{1,M})$; $\Xi^{s} \in
\mathbb{D}_{+}^{k_{s}} \ (s \in \overline{0,M})$, $\Upsilon_{0,s} \in
\mathbb{D}_{+}^{k_{s}} \ (s \in \overline{1,M})$;
$\{\Upsilon_{s,r}\}_{r=s+1}^{M} \subset \mathbb{D}_{+}^{n} \ (s \in
\overline{1,M-1})$; $\varrho\in\R$ and
$0<\Phi=\Phi^{\top}\in\mathbb{R}^{n\times n}$ such that
\begin{gather}
P + \varrho \sum_{j=1}^{\mu} \Lambda^j >0, \quad
Q=Q^{\top}=(Q_{a,\,b})_{a,\,b=1}^{M+2}\leq0, \nonumber \\\sum_{s=0}^{\phi} \Xi^s + 2\sum_{s=0}^{\phi} \sum_{r=s+1}^{\phi} \Upsilon_{s,r} >0, \nonumber
\end{gather}
where {\scriptsize
$Q_{1,1}=A_{0}^{\top}P +P A_{0} + \Xi^{0};\; \ Q_{j+1,j+1}= A_{j}^{\top} R_{j}^{\top}\Lambda^{j}+\Lambda^{j} R_{j} A_{j}+\Xi^{j},\;j\in\overline{1,M}; \;
Q_{1,j+1}=PA_{j}+ A_{0}^{\top} R_{j}^{\top}\Lambda^{j}+ R_{j}^{\top}\Upsilon_{0,j},\ j\in\overline{1,M};\; 
Q_{s+1,r+1}= A_{s}^{\top} R_{r}^{\top}\Lambda^{r}+\Lambda^{s} R_{s} A_{r}+R_{s}^\top R_{s}\Upsilon_{s,r} R_{r}^{\top} R_{r},\
s\in\overline{1,M-1}, r\in\overline{s+1,M};\;
Q_{1,\,M+2}=P,\; Q_{M+2,\,M+2}=-\Phi,\; 
Q_{j+1,\,M+2}=\Lambda^{j} R_{j},\;j\in\overline{1,M}$}.
Then, the system~\eqref{eq:change_variable_P_system} is ISS. 
\end{corollary}

The proof developments for Corollary~\ref{cor:extension_basis_function} refer to the methodology used in the works \cite{Efimov2019a} and are omitted from this paper due to the unobtrusive modifications.

\section{Numerical Case Studies} \label{sec:example}

\subsection{Preliminary Experiments}

This section deals with a system identification problem based on a simplified traffic flow dynamics model. The model incorporates both linear and nonlinear dynamics and considers the influence of external disturbances. We use the Generalized Persidskii System (GPS) method to identify this system and evaluate the identification effectiveness.

\subsubsection{Traffic Flow Model}

The traffic flow model under consideration can be written as:

\begin{equation} \label{eq:traffic_model}
\dot{x}(t) = Ax(t) + Bu(t) + f(x(t)) + d(t),
\end{equation}
where $x \in \mathbb{R}^2$ is the state vector, representing the traffic conditions (such as density or flow) of two adjacent road sections; $A \in \mathbb{R}^{2 \times 2}$ is the system matrix, describing the traffic flow relationship between road sections; $B \in \mathbb{R}^{2 \times 1}$ is the input matrix; $u(t) \in \mathbb{R}$ is a scalar input, which can represent traffic control measures; $f: \mathbb{R}^2 \rightarrow \mathbb{R}^2$ is a nonlinear function, representing the nonlinear effects of traffic states; $d(t) \in \mathbb{R}^2$ is external disturbance, which can represent various unforeseen factors.


In our simulation, we set the following parameters: The system matrix $A$ is a $2 \times 2$ matrix with elements $a_{11} = -0.1$, $a_{12} = a_{21} = 0.05$, and $a_{22} = -0.15$. The input matrix $B$ is a $2 \times 1$ column vector with elements $b_1 = 0.5$ and $b_2 = 0.3$. The nonlinear function $f(x)$ is defined as a two-dimensional vector-valued function, where the first component is $\sin(x_1)$ and the second component is $\cos(x_2)$, with $x_1$ and $x_2$ representing the first and second components of the state vector, respectively. The control input $u(t)$ is set to $\sin(t)$.

Here, the $\sin$ and $\cos$ functions are used to simulate the nonlinear effects of traffic states, which is a simplified approximation of complex traffic dynamics.

\subsubsection{Generalized Persidskii System Model}

We use a GPS model of the following form to identify the traffic flow system:

\begin{equation} \label{eq:gps_model}
\begin{bmatrix} \dot{x}(t) \\ \vdots \end{bmatrix}= \Gamma G(x(t), u(t)).
\end{equation}
In this model, we need to identify the parameter matrix \(\Gamma\), and use a vector \(G(x, u)\) composed of selected basis functions. We have chosen the basis functions satisfying Assumption~\ref{main_assum_sector_condition}, covering linear terms, cross terms, and some nonlinear functions. The control input basis function \(\psi(u)\) is $\begin{bmatrix} 0 \\ u \end{bmatrix}$ ($0$ is a zero column vector with appropriate dimension), containing only a simple linear term, and the state variable basis functions \(\varpi(x)\) include \(\varpi(x) = [x_1, x_2, x_1^3, x_2^3, \sinh(x_1), \sinh(x_2), \tanh(x_1), \tanh(x_2)]^\top\)

These basis functions are designed to capture both linear and nonlinear characteristics of the system, as well as the influence of the control input. The basis functions $\varphi(x)$ for the state variable $x$ contain, for example, linear terms and hyperbolic tangent functions, to capture various nonlinear features of the system. The basis function $\psi(u)$ for the control input $u$ is a linear term, representing the direct influence of the control input on the system.

The complete basis function vector $G(x, u)$ is constructed from the combination of these basic functions, also including a constant term 1 to capture the system's bias.

\subsubsection{Experimental Setup}

In the simulation, we set the time range as $t \in [0, 30]$, with the number of sampling points $N = 1000$, and initial state $x_0 = \begin{bmatrix} 1 & 3 \end{bmatrix}^\top$. To evaluate the model's performance under different disturbance conditions, we considered four different disturbance scenarios: Periodic disturbance $d_1(t) = 0.05 \begin{bmatrix} \sin(2t) & \cos(2t) \end{bmatrix}^\top$, random noise $d_2(t)$, which is Gaussian white noise with mean 0 and standard deviation 0.05, composite disturbance $d_3(t) = 0.1  \begin{bmatrix} \sin(t)\cos(2t) & \cos(t)\sin(2t) \end{bmatrix}^\top$, and large-amplitude disturbance $d_4(t) = \begin{bmatrix} \sin(2t) & \cos(2t) \end{bmatrix}^\top$. We first generated data from the system \eqref{eq:traffic_model} under these four disturbances, then used the EDMD method to identify the parameter matrix $\Gamma$ of the model \eqref{eq:gps_model}.

\subsubsection{Results Analysis}

We used Root Mean Square Error (RMSE) and Coefficient of Determination (R$^2$) to evaluate the model's performance. For the first three disturbance scenarios, the results are given as follows: For periodic disturbance, RMSE is 0.0489 and R$^2$ is 0.9954; for random noise, RMSE is 0.0059 and R$^2$ is 0.9999; for composite disturbance, RMSE is 0.0419 and R$^2$ is 0.9966.

In the fourth disturbance scenario, we found that the ISS condition in Theorem 1 could not be verified by solving the LMIs. This is an important observation, indicating that under large-amplitude disturbances, the system may lose its ISS property.

\begin{figure}[h!]
\centering
\includegraphics[width=0.8\textwidth]{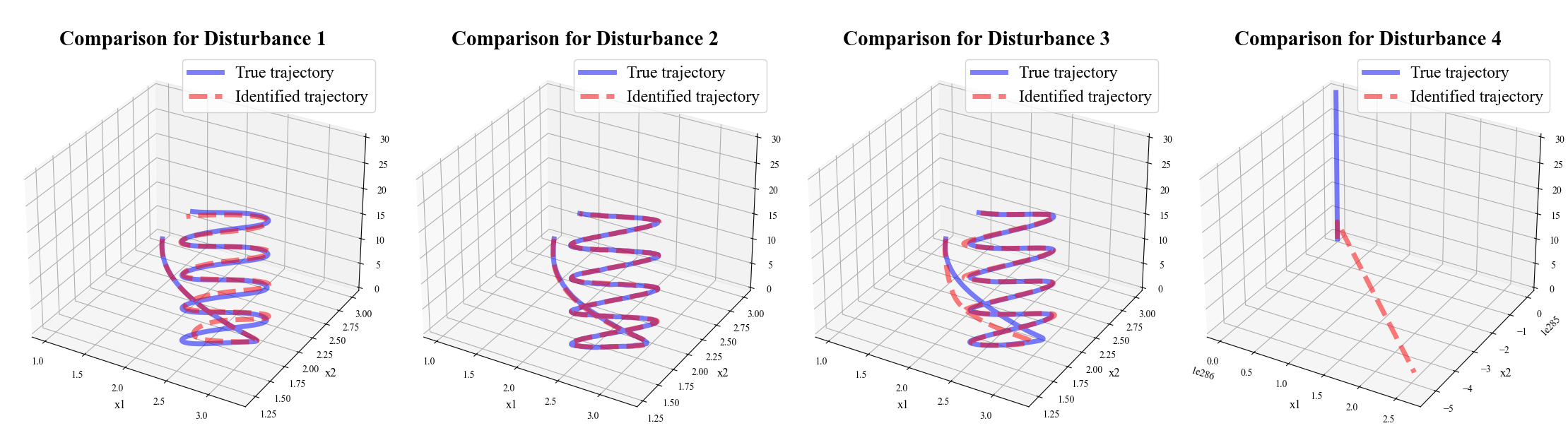}
\caption{Three-dimensional representation of true trajectories (blue solid lines) and identified trajectories (red dashed lines) under four disturbance scenarios.}
\label{fig:trajectories}
\end{figure}

Figure \ref{fig:trajectories} shows the three-dimensional representation of true trajectories and identified trajectories under four disturbance scenarios. Blue solid lines represent true trajectories, and red dashed lines represent identified trajectories.

These results indicate that the proposed GPS method can effectively identify and reconstruct the given traffic flow system, performing excellently in the first three disturbance scenarios. In these cases, the model demonstrates high accuracy, with R$^2$ values all exceeding 0.99, confirming the potential of this method in modeling and identifying complex dynamic systems.

However, for the fourth large-amplitude disturbance scenario, we were unable to verify the ISS condition through LMIs. This finding highlights the important relationship between system stability and disturbance amplitude. It suggests that when the disturbance amplitude exceeds a certain threshold, the system may lose its ISS property, which is significant for understanding and predicting the behavior of actual systems.

\subsection{Comparative Experiments on More Complex Systems} \label{subsec:compare_baseline}

To validate the effectiveness of our proposed GPS model, we designed a series of comparative experiments. We selected three commonly used nonlinear system modeling methods as baselines. These models underwent a comprehensive performance comparison with our GPS model.

\subsubsection{Setup of the System to be Identified}

This experiment employs a one-dimensional wave equation system with nonlinear terms, local external inputs, and global disturbances. This system can simulate physical phenomena such as vibrating strings, sound wave propagation, or disturbances in elastic media, providing an ideal testing platform for studying the identification and control of complex dynamical systems. The mathematical expression of the system can be given by
\[
\frac{\partial x}{\partial t} = c^2 \frac{\partial^2 x}{\partial y^2} + B(y)u(t) + f(x) + d(t).
\]
In this system, we set the following specific parameters and ranges:

The length of the spatial domain is set to $3\pi$ (approximately 9.42 meters), with a time span of 3 seconds. The spatial coordinate $y$ is in the range $[0, 3\pi]$, and time $t$ is in the range $[0, 3]$. The wave speed $c$ is set to 1 m/s. The input distribution function $B(y)$ takes the form of a square wave, being 1 in the middle 1/5 of the spatial domain and 0 elsewhere. The input signal $u(t)$ uses a composite sine wave, $u(t) = \sin(t) + 0.5 \cos(2t)$. The nonlinear function $f(x)$ is set as $f(x) = 0.5 \sin(x) + 0.3 \cos(2x) + 0.05x^2$. The external disturbance $d(t)$ is simulated as $d(t) = 0.1 \sin(3t)$.

The initial condition is set to $x(y,0) = \sin(y)$, and Neumann boundary conditions are adopted. We use the finite difference method to discretize the system spatially, with a spatial step size of approximately 0.094 meters, totaling 100 grid points. Temporally, we use numerical integration methods for ordinary differential equations, with a time step of 0.03 seconds, totaling 100 time steps.

During data collection, we record the values of the state variable $x$ at each spatial grid point. To evaluate the model's performance, we use RMSE and R$^2$ as evaluation metrics.

\subsubsection{Comparative Models}

To comprehensively evaluate the performance of our proposed GPS method, we selected the following three typical nonlinear system identification models as baselines:

\begin{enumerate}
\item \textbf{Autoregressive Exogenous Model (ARX)}: This is a classic linear system identification method. We used 2nd-order autoregressive and exogenous input terms. The advantage of the ARX model lies in its simplicity and computational efficiency, but it may not adequately capture complex nonlinear dynamics.
\item \textbf{Polynomial Regression Model (POLY)}: We used 2nd-order polynomial features, allowing the model to capture certain nonlinear relations. This model can be seen as an approximation of Taylor expansion, suitable for mild nonlinear systems.
\item \textbf{Radial Basis Function (RBF) Model}: The RBF model used 50 basis functions, capable of approximating various complex nonlinear functions. This model has powerful expressiveness but may require more training data and computational resources.
\end{enumerate}

All baseline models adopted Ridge regression as the basic regression method to prevent overfitting. In addition, we also applied preprocessing techniques such as feature selection and logarithmic scale transformation to improve the robustness and generalization ability of the models.

\subsubsection{Generalized Persidskii System Setup}
For our proposed Generalized Persidskii System, we selected the following basis functions: $\varpi(x) = [sinh(x), tanh(x), x, x^3]^\top$ and $\psi(u) = u^\top$.

\subsubsection{Experimental Results and Analysis}
The experimental results are shown in the following table:

\begin{table}[htbp]
\centering
\begin{tabular}{|l|c|c|}
\hline
Model & RMSE & R$^2$ \\
\hline
ARX Model & 1.0329 & -5.0861 \\
POLY Model & 0.7528 & -2.7528 \\
RBF Model & 0.7489 & -2.7567 \\
GPS (ours) & 0.1125 & 0.9163 \\
\hline
\end{tabular}
\caption{Performance Comparison of Different Models}
\label{tab:model_comparison}
\end{table}

From the results in Table \ref{tab:model_comparison}, it can be seen that the GPS significantly outperforms other baseline models on all evaluation metrics. Its RMSE is the lowest (0.1139), only about 15\% of the other models, while its R$^2$ value is 0.9488, close to 1, indicating that the model explains most of the variation in the original data. In contrast, the ARX model performs the worst, with the highest RMSE (1.0329) and a negative R$^2$ value (-5.0861), indicating that the linear model is completely unsuitable for describing the strongly nonlinear system. The polynomial regression model and RBF model perform similarly, both showing improvement over the ARX model, but still far behind the GPS. Their RMSEs are around 0.75, and R$^2$ values are still negative (about -2.75). The excellent performance of the GPS may be attributed to its structure being closer to the dynamic characteristics of the original system. It not only captures nonlinear relations but also effectively describes the temporal evolution characteristics of the system.

\begin{figure}[h!]
\centering
\includegraphics[width=1\textwidth]{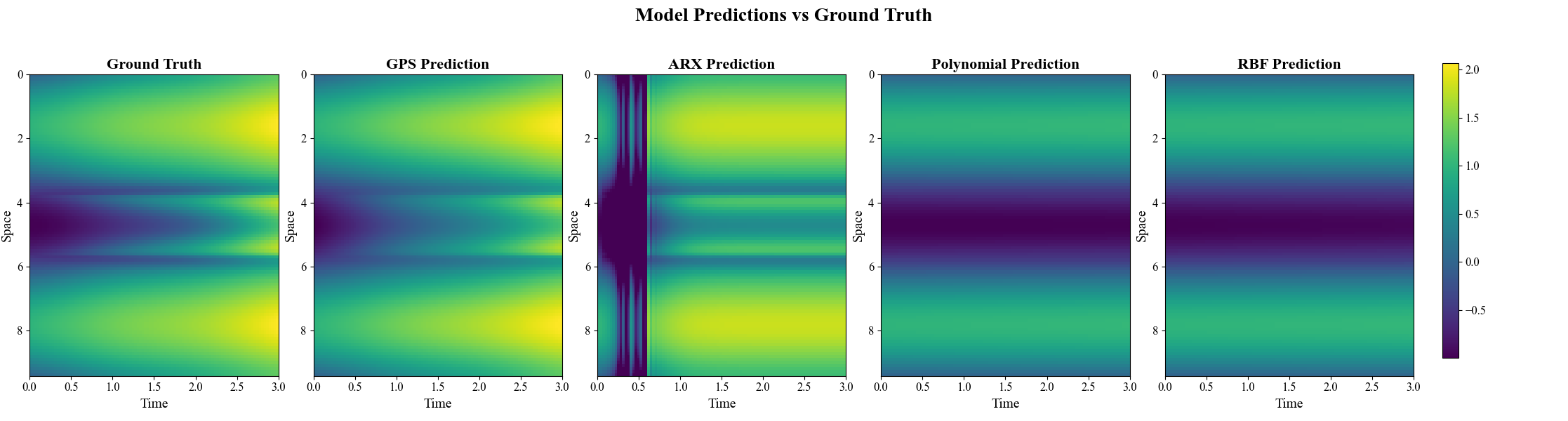}
\caption{Comparison of Prediction Results from Different Models with Real Data}
\label{fig:model_comparison}
\end{figure}

Figure \ref{fig:model_comparison} shows the comparison between the prediction results of each model and the real data. From the figure, it can be intuitively seen that the predicted trajectory of the GPS (our proposed method) is closest to the real data, while other models show larger deviations. This visual comparison further confirms the conclusions we previously drew based on numerical metrics.

\subsection{Comparative Experiments with Different Basis Function Combinations}

This section explores the impact of different basis function combinations on the identification effectiveness of the GPS model. We use the same traffic flow model as in the previous section, focusing on how the choice of basis functions affects model performance.

\subsubsection{Experimental Setup}

This experiment employs a simplified one-dimensional wave equation system, similar to the previous experiment in Section~\ref{subsec:compare_baseline}. The system dynamics equation can be modeled by
\begin{equation}
\frac{\partial x}{\partial t} = Ax + Bu(t) + f(x,t) + d(t),
\end{equation}
where $A$ is the discretization matrix of the second-order spatial derivative, $B$ is the input matrix effective only in the middle fifth of the space. The nonlinear term $f(x,t) = \sin(\omega t) \sin(x) + 0.5 \cos(2\omega t) \cos(2x)$ depends on both space and time. The external input $u(t) = \sin(t) + 0.5\cos(2t)$, and $d(t)$ is the disturbance.

The spatial domain is set to $[0, \pi]$, the time range is $[0, 2]$, each using 300 discrete points. The initial condition is $x_0 = \sin(x) - 0.7$, with Dirichlet boundary conditions. The wave speed $c = 1$, spatial step size $\Delta x = 1$, and the angular frequency of the periodic component $\omega = 0.5\pi$.

We consider the following four basis function combinations: Combination 1 (Simple) is $\varpi_1(x) = \begin{bmatrix} \sinh(x) & \tanh(x) & u^\top \end{bmatrix}^\top$, Combination 2 (Adding linear term) is $\varpi_2(x) = \begin{bmatrix} \sinh(x) & \tanh(x) &  x & u^\top \end{bmatrix}^\top$, Combination 3 (Adding quadratic term) is $\varpi_3(x) = \begin{bmatrix} \sinh(x) & \tanh(x) & x & x^3 & u^\top \end{bmatrix}^\top$, and Combination 4 (Complex) is $\varpi_4(x) = \begin{bmatrix} 1 & \sinh(x) & x & x^3 & x^5 & u^\top \end{bmatrix}^\top$.

We use composite disturbances to test the robustness of the model: $d(t) = 0.1 \begin{bmatrix} \sin(t)\cos(2t) & \cos(t)\sin(2t) \end{bmatrix}^\top$.

\subsubsection{Result Analysis}

We use RMSE and R$^2$ to evaluate the performance of each basis function combination. The results are shown in Table~\ref{tab:bfs_comparison}.

\begin{table}[h!]
\centering
\begin{tabular}{|c|c|c|}
\hline
Combination & RMSE & R$^2$ \\
\hline
Combination 1 & 0.0086 & 0.9998 \\
Combination 2 & 0.0067 & 0.9999 \\
Combination 3 & 0.0043 & 0.9999 \\
Combination 4 & 0.0021 & 1.0000 \\
\hline
\end{tabular}
\caption{Performance Comparison of Different Basis Function Combinations}
\label{tab:bfs_comparison}
\end{table}
According to the results in Table~\ref{tab:bfs_comparison}, all basis function combinations demonstrate high accuracy, with R$^2$ values exceeding $0.9998$. As the complexity of basis functions increases, model performance shows a gradual improvement trend, with Combination 4 achieving the lowest RMSE (0.0021) and the highest R$^2$ value (1.0000). However, the magnitude of performance improvement decreases with increasing complexity, as seen in the relatively small improvement from Combination 2 to Combination 3. This indicates that the GPS model can effectively capture system dynamics under various basis function selections, but in practical applications, it is necessary to balance model complexity and computational cost to choose an appropriate basis function combination.


\section{Conclusion}
This work proposes a beneficial class of functions named \emph{Sector Basis Functions} for identifying the vector field of the system. The identified system takes the form of a specific type of nonlinear system whose input-to-state stability conditions were formulated as linear algebraic inequalities and, thus, can be constructively verified. Additionally, two directions of the extensions of the basis functions are introduced. Simulation of the macroscopic traffic dynamics and a more complex one-dimensional wave equation system with nonlinear terms were exploited for examining the proposed results. These simulations demonstrated the effectiveness of our approach in identifying and reconstructing complex dynamical systems under various disturbance scenarios.

\paragraph{\textbf{Limitations and Future Work}} Although our method shows promise, there are several limitations and areas for future research. First, the performance of the method under large amplitude disturbances requires further investigation, as we observed challenges in verifying the ISS condition in such scenarios. Second, the selection of appropriate basis functions for different types of systems remains a challenge and could benefit from further study.

Future work topics include further applications to power system dynamics and extending the theory to descriptor systems. The application to power system dynamics could provide valuable insights into grid stability and control under varying load conditions and renewable energy integration. Extending the theory to descriptor systems would broaden the applicability of our method to a wider class of systems, including those with algebraic constraints.

Building upon our current research, we will explore methods to relax specific mathematical constraints in our framework. Our focus will be on relaxing the strict sector condition $\nu f_{j'}^{i}(\nu) > 0$ for all $\nu \neq 0$. To this end, we will investigate the application of machine learning techniques to adaptively learn and approximate nonlinear functions that may violate these conditions. We aim to develop a dynamic framework combining our theoretical framework with data-driven methods, using neural networks or other flexible function approximators to model system components that exhibit different characteristics. This approach could extend our method to a wider class of nonlinear systems while preserving analytical traceability.

Additionally, future research could explore adaptive methods for selecting and refining basis functions based on system behavior, potentially improving the model's accuracy and robustness. The integration of machine learning techniques, such as deep learning, with our GPS method could also be an interesting avenue for enhancing the identification of highly complex nonlinear systems.

Finally, addressing implementation-level issues in Koopman operator theory, particularly the challenge of dealing with cases where the logarithm of the linear representation matrix may not exist due to negative eigenvalues, remains an important area for future investigation. This could involve developing alternative representations or numerical methods to handle such cases effectively.

\section*{Declaration of Competing Interest} The authors declare that they have no known competing financial interests or personal relationships that could have appeared to influence the work reported in this paper.

\section*{Acknowledgments} This work is supported by the National Science Foundation (NSF) under grants 2152450 and 2151571. It is also supported by the Natural Science Foundation of Jiangsu Province, and the Fundamental Research Funds for the Central Universities under grants 2242024k30037 and 2242024k30038. 


\section*{CRediT Authorship Contribution Statement}
\textbf{Wenjie Mei:} Conceptualization, Methodology, Investigation, Writing - original draft, Supervision, Funding acquisition.
\textbf{Dongzhe Zheng:} Software, Validation,  Writing - review \& editing, Visualization, Investigation.
\textbf{Yu Zhou:} Investigation, Writing - review \& editing. \textbf{Ahmad Taha:}  Funding acquisition, Writing - review \& editing. \textbf{Chengyan Zhao:} Writing - review \& editing.

\appendix
\section*{Appendices} 
\section{General dynamical systems} \label{appendix:general_system}
This section gives the used preliminaries for the general continuous-time nonlinear dynamical system:
\begin{gather}\label{eq:general_nonlinear_system}
\dot{x}(t) =F(x(t),u(t)),\quad  t \in \mathbb{R}_{+},
\end{gather} 
where $x(t) \in\mathbb{R}^{n}$ is the state vector; $u(t) \in \mathrm{U} \subset \mathbb{R}^{m}$
is the given external input, $u\in C^{1}_{m}([0,\infty))$; and vector-valued nonlinearity is defined as $F \in C(\mathbb{R}^{n}\times \mathrm{U}, \mathbb{R}^{n})$ being also locally Lipschitz continuous in $x(t)$. For an initial state $x_{0}\in\mathbb{R}^{n}$, $u\in C^{1}_{m}([0,\infty))$ and $t \in\mathbb{R}_{+}$, the corresponding solution
of system~\eqref{eq:general_nonlinear_system} is denoted by $x^t(x_{0},u) = x(t,x_0,u)$. It is assumed that in~\eqref{eq:general_nonlinear_system} such a solution is uniquely defined for any $x_0\in\R^n,  u\in C^{1}_{m}([0,\infty))$ and all $t\in\R_{+}$.

\section{Basis of Koopman operator}
We present the definition of the Koopman operator \cite{Mauroy2020} $K^t: \mathcal{H} \rightarrow \mathcal{H}$ ($t\in \mathbb{R}_{+}$) associated with the system~\eqref{eq:general_nonlinear_system} as 
\begin{equation*}
K^t H = H  \circ  x^t, 
\end{equation*}
where $H \colon \mathbb{R}^n  \rightarrow \mathbb{R}$ is observable function ($H$ belongs to a Banach space $\mathcal{H}$ of such functions) and $\circ$ denotes the composition operator. The Koopman operator is of interest since it can transform nonlinear dynamics into a linear representation of the theoretically infinite-dimensional system.  Let $\mathfrak{K} = \{K^t\}_{t \geq 0}$ represent the $C_0$-semigroup of Koopman operators $K^t$, then $L := \lim_{t \to 0^{+}} \frac{K^t - I}{t}$ stands for the infinitesimal generator of $\mathfrak{K}$, where $I$ is the identity operator on the space $\mathcal{H}$. Note that theoretically, $\mathcal{H}$ can be infinite-dimensional, and $K^t$ is a linear operator for each $t\in \R_{+}$.

\section{Input-to-state stability properties}
%
\begin{defn}  {\em \cite{sontag1996new,Sontag1995}}  
\label{def:IOS} A forward complete system~\eqref{eq:general_nonlinear_system}
is said to be \emph{input-to-state stable (ISS)} if there exist $\beta\in\mathscr{KL}$,
$\gamma\in\mathscr{K}$ such that 
$
\rVert x^t(x_{0},u)\rVert\leq\beta\left(\rVert x_{0}\rVert,t\right)+\gamma(\rVert u\rVert_{\infty}),\ \forall t\in\mathbb{R_{+}}
$
for any $x_{0}\in\mathbb{R}^{n}$ and $u\in C^{1}_{m}([0,\infty))$.
For the system~\eqref{eq:general_nonlinear_system}, a smooth function $V\colon\mathbb{R}^{n}\rightarrow\mathbb{R}_{+}$
is an {\em ISS-Lyapunov function} if there exist $\alpha_{1},\alpha_{2},\alpha_{3}\in\mathscr{K}_{\infty}$
and $\chi\in\mathscr{K}$ such that 
\begin{gather} \label{eq:ISS_V}
\alpha_{1}(\rVert x\rVert)\leq V(x)\leq\alpha_{2}(\rVert x\rVert),\\
\rVert x\rVert\geq\chi(\Vert u\Vert) \quad  \Rightarrow \quad \nabla V(x)F(x,u)\leq-\alpha_{3}(\rVert x\rVert)\nonumber 
\end{gather}
for all $x\in\mathbb{R}^{n}$ and $u\in \mathrm{U}$.
\end{defn}


\begin{thm}{\em \cite{sontag1996new,Sontag1995}} 
The system~\eqref{eq:extended_general_nonlinear_system} is ISS if and only if it admits an ISS-Lyapunov function. \label{thm:main_ISS_Ly}
\end{thm}


\section{Generalized Persidskii systems} \label{appendix:gps}
We then introduce \emph{generalized Persidskii dynamics} \cite{mei2022nonlinear,Efimov2019a}:
\begin{equation}
\dot{x}(t)=A_{0}x(t)+\sum_{j'=1}^{M}A_{j'}F_{j'}(x(t))+u(t),\quad t\in\mathbb{R}_{+},\label{eq:main_system_Per}
\end{equation}
where $x=[x_{1} \ldots x_{n}]^{\top}\in\mathbb{R}^{n}$ is the state, $x(0)=x_{0}$; $A_s \in \R^{n \times n} $, $s\in\overline{0,M}$ are constant matrices; the input $u=[u_{1}, \ldots, u_{n}]^{\top}\in\mathscr{L}_{\infty}^{n}$ (the space of functions $u$ with
        $\rVert u\rVert_{\infty}:=\rVert u\rVert_{[0,\infty)} = \text{ess}\sup_{t\in[0, \infty)}\rVert
      u(t)\rVert <+\infty$), and the functions
$
F_{j'} \in C(\mathbb{R}^{n}, \mathbb{R}^{n}), \;
F_{j'}(x)=[\begin{array}{ccc}
f_{j'}^{1}(x_{1}) & \ldots & f_{j'}^{n}(x_{n})\end{array}]^{\top},\;\forall j'\in\overline{1,M}.
$
The nonlinearity $F_{j'}$ has a diagonal structure: Each element of $F_{j'}$ (\textit{i.e.}, each function $f_{j'}^{i}$) depends only on the respective coordinate $x_{i}$, $i \in\overline{1,n}$.  We also impose a sector boundedness or a passivity condition on $f_{j'}^{i}$:
\begin{assum} \label{main_assum_sector_condition}
For system~\eqref{eq:main_system_Per}, assume that for any $i\in \overline{1,n}$ and $j' \in \overline{1,M}$, we have 
$
\nu f_{j'}^{i}(\nu)>0$ for all $\nu \neq 0.
$
\end{assum}

Under Assumption~\ref{main_assum_sector_condition}, with a reordering of nonlinearities and their
decomposition, there exists an index $\phi \in\overline{0,M}$ such that
for all $a\in\overline{1,\phi}$, $i\in\overline{1,n}$: 
$\lim_{\nu\rightarrow\pm\infty}f_{a}^{i}(\nu)=\pm\infty$,
and that there exists $\mu\in\overline{\phi,M}$ such that for all 
$b\in\overline{1,\mu}$, $i\in\overline{1,n}$: 
$
\lim_{\nu\rightarrow\pm\infty}\int_{0}^{\nu}f_{b}^{i}(\tau)d \tau=+\infty.
$ \label{main_assum_foot} 

Assumption~\ref{main_assum_sector_condition} is not restrictive and often appears in the field of deep learning, considering, for instance, the identity rectified linear unit (ReLU), $\tanh$, and sigmoid functions after potential translation are some representative examples fulfilling this assumption.

 \bibliographystyle{elsarticle-num} 
 \bibliography{ieeeconf}





\end{document}